# Direct observation of ultrafast thermal and non-thermal lattice deformation of polycrystalline Aluminum film

Runze Li[1,2*], Pengfei Zhu[1], Jie Chen[1], Jianming Cao[1,3], Peter M. Rentzepis[2] and Jie Zhang[1**]

[1]*Key Laboratory for Laser Plasmas (Ministry of Education), School of Physics and Astronomy, and Collaborative Innovation Center of Inertial Fusion Sciences and Applications (CICIFSA), Shanghai Jiao Tong University, Shanghai 200240, China*

[2]*Department of Electrical and Computer Engineering, Texas A&M University, College Station, TX 77843, USA*

[3]*Physics Department and National High Magnetic Field Laboratory, Florida State University, Tallahassee, Florida 32310, USA*

*\* Email: Runze@ece.tamu.edu; \* \*Email: jzhang1@sjtu.edu.cn*

**Abstract:** The dynamics of thermal and non-thermal lattice deformation of nanometer thick polycrystalline aluminum film has been studied by means of femtosecond (fs) time-resolved electron diffraction. We utilized two different pump wavelengths: 800 nm, the fundamental of Ti:sapphire laser and 1250 nm generated by a home-made optical parametric amplifier (OPA). Our data show that, although coherent phonons were generated under both conditions, the diffraction intensity decayed with the characteristic time of 0.9±0.3 ps and 1.7±0.3 ps under 800 nm and 1250 nm excitation, respectively. Because the 800 nm laser excitation corresponds to the strong interband transition of aluminum due to the 1.55 eV parallel band structure, our experimental data indicate the presence of non-thermal lattice deformation under 800 nm excitation, which occurs at a time-scale that is shorter than the thermal processes dominated by electron-phonon coupling under 1250 nm excitation.

The ultrafast structural deformation of metallic crystals is not only fundamentally important but also crucial for precision fabrication of materials by femtosecond laser pulses. With the development of ultrafast techniques in the past few decades, the non-equilibrium processes of materials, under femtosecond laser excitation, has been widely studied by means of ultrafast optical [1, 2], time-resolved x-ray and electron diffraction methods [3-8]. The ultrafast structural deformation and phase transition mechanisms of metals under femtosecond laser irradiation may be mainly classified as thermal, non-thermal or combination of the two. For example, the thermal and non-thermal deformation of tungsten irradiated with femtosecond laser pulses have been revealed by time-resolved transient reflectivity measurements[9].

Upon femtosecond laser irradiation, the photon energy is initially deposited into the electron system through photon-electron interaction. Because the electron heat capacity is typically orders of magnitude smaller than that of the lattice, the electron system is therefore heated to a very high temperature and equilibrates through electron-electron interaction while the lattice system remains, "cold", at room temperature. Such a non-equilibrium system will reach a new equilibrium state within picoseconds, which is dominated by the electron-phonon and phonon-lattice interactions [10]. For such thermal mechanism, with sufficient excitation energy, the lattice temperature may exceed the melting point and the collapse of lattice order will occur. It has been shown by ultrafast time-resolved x-ray diffraction studies that the melting of metals such as gold, silver and copper is a purely thermal process under moderate femtosecond laser excitation [11, 12].

The non-thermal mechanism, however, typically involves electronic excitation that directly leads to lattice instability. For example, photon excitation of electrons may alter directly the interatomic potential and cause lattice disorder or melting [13]. Because such electronic excitation occurs faster than the electron thermalization and electron-phonon coupling process, the 1/e characteristic time for such non-thermal deformation is typically on the order of sub-picosecond time interval and therefore much faster than the picoseconds time thermal mechanism. In addition, the threshold of non-thermal melting is also lower than that of thermal melting [14]. For example, ultrafast x-ray diffraction experiments reveal that the melting of Germanium is faster than the time-scale set by the propagation of supersonic melting front, indicating a non-thermal process [15].

Despite the numerous experimental and theoretical studies of metals, the lattice deformation mechanism of aluminum irradiated with femtosecond laser pulses is still under debate. An earlier ultrafast transient reflectivity measurement has shown that the dielectric constant of solid aluminum measured within hundreds of femtoseconds after 800 nm laser irradiation, is equal to its liquid state dielectric constant[14]. Since this time scale is much shorter than the picosecond lattice thermalization, it suggests that the dominant process is non-thermal. However, a later transient reflectivity study, which employed broadband light source, claimed that the melting of aluminum around 800 nm excitation does not involve non-thermal component [16]. Meanwhile, with the improvement of electron pulse width from picoseconds to sub-picosecond, the observed melting time measurement of aluminum crystal is decreased from 20 ps to 3.5 ps, both implying a time constant that is due to thermal disorder [8, 17]. However, the observed diffuse scattering intensity of ultrafast electron diffraction study supported the existence of a non-thermal process for aluminum irradiated with 800 nm laser pulses[18]. In the present study, we employed two different laser wavelengths for excitation, 800 nm and 1250 nm, which allow us to compare the lattice deformation process of aluminum, with and without the interband transition of the electronic systems due to the 1.55 eV parallel energy structure. Our results reveal that, the characteristic time of lattice deformation under 1250 nm irradiation is about twice as long as that under 800 nm excitation. Therefore, these experiments provide direct evidence that the heating of aluminum lattice under 800 nm excitation involves non-thermal components while the process under 1250 nm illumination is a thermal process.

The ultrafast electron diffraction experimental system employed in this study has been described in detail previously [19]. Briefly, the 800 nm, 70 fs, 1 mJ, laser pulse from a 1 KHz Ti:sapphire laser system is split into two parts. 40% of the laser beam is frequency tripled to 266 nm and directed to a photocathode to generate femtosecond electron pulses, which are accelerated to 59 keV and used to probe the 25 nm thick polycrystalline aluminum sample in a transmission diffraction configuration. The remaining 60% of the laser beam, which is used to excite the sample, is directed onto a translation delay stage that can precisely control the delay time between the arrival of the pump laser pulse and the probe electron pulse at the sample. Wavelength tunability, and the 1250 nm pump pulse, are generated through a home-made Optical Parametric Amplifier (OPA) [20-22], which is illustrated in Figure 1 and integrated into the pump beam path of our femtosecond electron diffraction system.

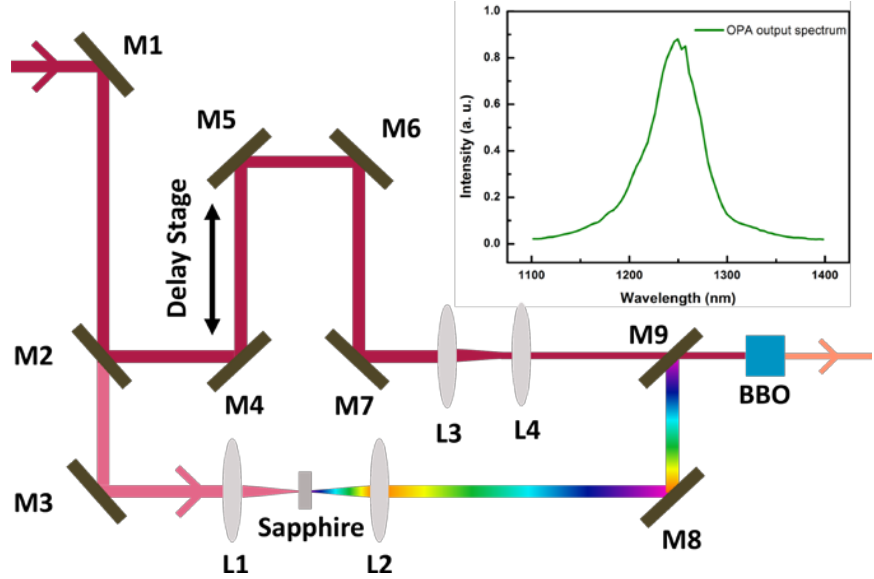

Figure 1: Schematic representation of the home-made optical parametric amplifier (OPA) used in this experiment. The OPA is capable of generating 1250 nm femtosecond laser pulses and its output spectrum is shown in the insert, which indicates a full-width at half maximum (FWMH) of ~ 55 nm. The OPA is integrated into the pump beam path and the in-suit switching between 800 nm and 1250 nm irradiation of the Al sample is realized by removing or inserting the 5-mm thick BBO crystal and filters, which has negligible effects on the pump beam path.

This home-made OPA employed a type II phase match: 800 nm (e) = 1250 nm (o) + 2300 nm (e) and the 5-mm thick Beta-Barium Borate (BBO) crystal was cut to $\theta = 29.3°$, $\phi = 30°$ (CASTECH Inc.). The white light continuum (WLC), which was due to the interplay of self-focusing and self-phase modulations and served as the signal pulse, was generated by focusing ~3 μJ/pulse 800 nm laser onto a 3 mm thick sapphire plate. The majority of the 800 nm pump laser was collimated onto the BBO crystal by a lens pair, L3 and L4, with an irradiation intensity of ~ 40 GW/cm$^2$. The temporal overlap between the 800 nm pump laser pulses and the signal pulses is achieved by adjusting the delay stage in order to maximize the amplification. Bandpass filters and 800 nm total reflection mirrors were inserted to the output beam of the OPA to filter out the residual 800 nm pump laser pulses and idle pluses. The fluence of the 1250 nm pulse was adjusted according to the 800 nm experiments to make certain that the energy absorbed by the Al film is the same for both pump wavelengths. In the 800 nm experiments, the pump laser was focused onto the 25 nm thick Al sample with a fluence of 2.1 mJ/cm$^2$, which was far below the damage threshold of our Al sample. Except for the pump laser wavelength, all other factors such as the number of electrons per probe pulse, the optical irradiation area, the electron probe area on the Al sample, the laser energy absorbed by the sample, and the integration time for the diffraction patterns are the same for both sets of experiments that utilize 800 nm and 1250 nm pump pulses.

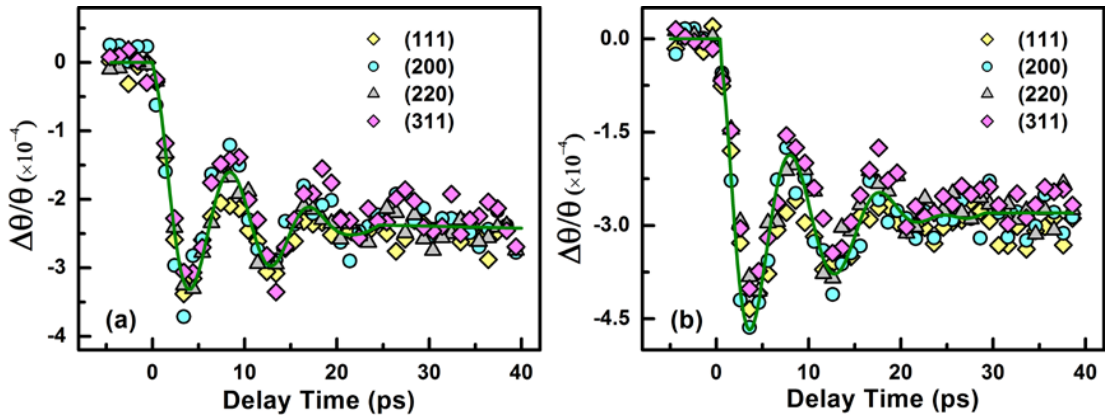

Figure 2: Time-dependent relative change of the diffraction angle under (a) 800 nm and (b) 1250 nm light pulse excitation.

The experimental data are presented in the form of the relative shift of the diffraction peak positions and the changes in the integrated diffraction intensities. According to the first-order Bragg diffraction formula, $2d\sin\theta = \lambda$, and the small angle approximation, we have $\Delta\theta/\theta \approx \Delta\theta/\tan\theta = -\Delta d/d$, where $d$, $\theta$, and $\lambda$ are the lattice plane distance, Bragg diffraction angle, and the wavelength of the probe electrons, respectively. Therefore, the relative changes of the diffraction angles, $\Delta\theta/\theta$, represents the relative change of lattice plane distance, $\Delta d/d$. It is shown in Figure 2 that, for both 800 nm and 1250 nm excitation wavelengths, the relative change in lattice plane distance experienced several damping oscillations after time zero, before it stabilized at a new equilibrium position at ~ 25 ps after laser irradiation. The damping oscillations indicate the generation of coherent phonons [23, 24], which represent the breathing motion of the lattice along the sample surface normal direction, generated through the induced strain and stress [25]. Similar oscillations have been observed in various ultrafast time-resolved diffraction studies of single crystals such as Gold, Silver, Copper and polycrystalline Aluminum [11, 12, 26, 27]. The oscillations indicate the formation of a standing wave between the two surfaces of the Al sample. Given the sound velocity of solid Al, $V$ =6420 m/s, and the film thickness, $L$ = 25±3 nm, the oscillation period is, $T = 2L/V$ =7.8±0.9 ps, which agrees well with the experimentally observed period of 8.2 ps. The damping process of the oscillations is due to the energy that has been dissipated into the surroundings. As expected, for both the 800 nm and 1250 nm excitations, the oscillation period of the time-dependent relative change of the lattice plane distance is the same within the experimental time resolution. Most important, the newly established equilibrium positions around 30 ps after laser irradiation are $2.39\times10^{-4}$ and $2.82\times10^{-4}$ for the 800 nm and 1250 nm excitation, respectively. This indicated that the light energy actually deposited into the Al sample are essentially the same for both excitation wavelengths, commensurate with the design and expectations of this experiment.

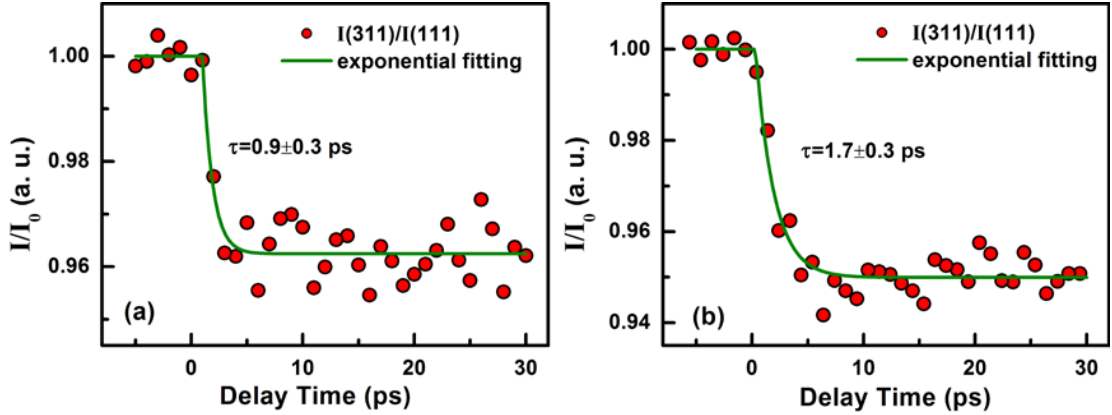

Figure 3: Time-dependent evolution of the normalized total diffraction intensity under (a) 800 nm and (b) 1250 nm excitation.

Table 1: Summary of the main results under 800 nm and 1250 nm light excitations

| Pump Wavelength | Equilibrium $\Delta\theta/\theta$ ($\times 10^{-4}$) | Max. $I/I_0$ (a.u.) | $\tau_I$ (ps) |
|---|---|---|---|
| 800 nm | 2.39 | 96.3% | 0.9±0.3 |
| 1250 nm | 2.82 | 95.0% | 1.7±0.3 |

The time-dependent electron diffraction intensities under 800 nm and 1250 nm excitation are shown in Figure 3 and their comparison with the lattice plane distance change is depicted in Table 1. According to Debye-Waller effects, the diffraction intensity represents the mean-square-displacement of lattice atoms, which is the deviation with respect to the lattice position of a perfect crystal. It is shown in Figure 3 and Table 1 that the maximum intensity drop under the two excitation wavelengths is similar, further indicating that the light energy deposited into the Al sample are practically the same. However, the exponential fitting of the time-dependent evolution of the diffraction intensity indicates that the characteristic time of the diffraction intensity drop is 0.9±0.3 ps and 1.7±0.3 ps, for 800 nm and 1250 nm light excitation, respectively. The ~0.9 ps characteristic time constant agrees well with previous ultrafast electron diffraction studies of Al crystal irradiated with 800 nm laser pulse [27]. The faster energy relaxation rate under 800 nm excitation may be explained by the non-thermal deformation mechanism due to the parallel-band structure of Al. It has been reported that, because of the parallel-band structure in planes parallel to the (200) faces of the Brillouin zone of face-centered cubic Aluminum, there is a strong interband transition around 1.55 eV, which corresponds to the photon energy of 800 nm excitation [14, 28-30]. Such 800 nm laser excitation has been found to alter the dielectric constant at a threshold fluence and a time scale that is less than the typical values predicted by ultrafast thermal heating, indicating that a non-thermal mechanism is possibly responsible for the collapse of the lattice order [14]. Because such non-thermal (electronic) mechanism does not involve the electron-phonon coupling process but alters the lattice structure directly through electronic excitation, the time scale of structural deformation dominated by the non-thermal mechanism is shorter than that of the thermal mechanism. Therefore, under the 1250 nm light excitation, which is lower than the parallel band gap of Al and does not involve electronic excitation, we propose that the Al sample experiences a purely thermal deformation that is governed by electron-phonon coupling and represents a longer time than that with non-thermal 800 nm electronic excitation, 1.7 ps vs 0.9 ps. Our experimental results also agree with previous ultrafast time-resolved diffuse scattering studies: it has been shown that, due to the non-thermal deformation mechanism of Al crystal irradiated with 800 nm laser pulses, the characteristic time of short-range disorder is smaller than that of the long-range disorder [18]. Given that the diffraction intensity change is determined by both the long-range and short range disorders, the convolution of the two types of disorder may reveal a shorter time scale under 800 nm laser irradiation, when compared to the same amount of laser energy deposited into the lattice by the 1250 nm laser pulses through a purely thermal process. In addition to Al, non-thermal deformation of lattice is also possible in other metals and semiconductors under femtosecond laser excitation. Those include a non-thermal component for the sub-picosecond destruction of tungsten due to the Fermi redistribution within the d-band pseudogap [9], and other ultrafast non-thermal phase-transition in germanium[15], GeSb film [31] , and silicon [32].

In summary, a home-made optical parametric amplifier, which is capable of generating 1250 nm femtosecond laser pulses, was incorporated into our ultrafast time-resolved electron diffraction system. The structural dynamics of a 25 nm thick polycrystalline Aluminum film under 800 nm and 1250 nm laser excitation was studied by means of ultrafast time-resolved transmission diffraction configuration. The experimental data indicate that, under practically the same laser energy deposition into the sample, the 0.9±0.3 ps characteristic time in the diffraction intensity drop under 800 nm excitation, is much shorter than the 1.7±0.3 ps measured under 1250 nm excitation. The 800 nm excitation photons, which correspond to the 1.55 eV parallel band structure of Al, induce a strong interband transition and may cause a non-thermal deformation of the lattice. Such electronic excitation lead to a shorter time constant for the lattice deformation. This study provides experimental evidence that the deformation of Al lattice under 800 nm excitation involves a non-thermal process, while the 1250 nm excitation induces a purely thermal process that is governed by electron-phonon coupling.

**Acknowledgements:**

This research was supported in part by the National Natural Science Foundation of China (Grant Numbers: 11674227, 11327902 and 11421064), the Welch Foundation (Grant Number: 1501928) and Texas A&M University TEES.